\documentclass[9pt,twocolumn,twoside]{osajnl}
\journal{ol} 

\setboolean{shortarticle}{true}

\title{Investigation of $Q$ degradation in low-loss Si$_3$N$_4$ from heterogeneous laser integration}

\author[1,*]{Joel~Guo}
\author[1]{Chao~Xiang}
\author[1]{Warren~Jin}
\author[1]{Jonathan~Peters}
\author[1]{Mingxiao Li}
\author[1]{Theodore Morin}
\author[1]{Yu Xia}
\author[1]{John~E.~Bowers}

\affil[1]{Department of Electrical and Computer Engineering, University of California, Santa Barbara, Santa Barbara, California 93106, USA}

\affil[*]{Corresponding author: joelguo@ucsb.edu}

\begin{abstract} 
High-performance, high-volume-manufacturing Si$_3$N$_4$ photonics requires extremely low waveguide losses augmented with heterogeneously integrated lasers for applications beyond traditional markets of high-capacity interconnects. State-of-the-art quality factors ($Q$) over 200 million at 1550~nm have been shown previously; however, maintaining high $Q$s throughout laser fabrication has not been shown. Here, Si$_3$N$_4$ resonator intrinsic $Q$s over 100 million are demonstrated on a fully integrated heterogeneous laser platform. $Q_i$ is measured throughout laser processing steps, showing degradation down to 50 million from dry etching, metal evaporation, and ion implant steps, and controllable recovery to over 100 million from annealing at 250~$^\circ C$ - 350~$^\circ C$.
\end{abstract}

\setboolean{displaycopyright}{true}

\begin{document}

\maketitle

Silicon nitride (Si$_3$N$_4$) photonics has developed rapidly in the last decade, leveraging CMOS-compatibility as well as favorable material properties such as a high effective $\chi^{(3)}$ nonlinearity, a wide transparency window, and low optical losses~\cite{xiang2022silicon}. These properties allow for powerful functionalities such as frequency combs~\cite{ji2024multimodality}, photogalvanic-induced second harmonic generation~\cite{li2023high,billat2017large}, magneto-optical trap beam delivery~\cite{isichenko2023photonic}, feedback tolerant lasers~\cite{white2024unified,xiang20233d}, low-frequency noise rivaling fiber lasers~\cite{li2021reaching}, and tunable lasers beyond the silicon bandgap~\cite{tran2022extending,zhang2023photonic,zhang2023integrated}. Notably, all of these functions depend heavily on ultra-low losses, which can be equivalently represented by the quality factor, $Q$. 

Despite the powerful functionalities available in Si$_3$N$_4$, heterogeneous laser integration is still required to scale in terms of economics or circuit complexity. As photonic systems grow in complexity, electrically-pumped optical gain on the same chip is necessary to compensate losses while allowing for denser integration~\cite{xiang2021perspective}; similarly in quantum computing systems, individually addressing millions of atoms or ions necessitates dense integration of lasers on chip~\cite{moody2021roadmap,niffenegger2020integrated}. In terms of economic viability, Intel demonstrated commercial deployment of millions of 800 Gbps optical transcievers, leveraging heterogeneous laser integration via wafer bonding on a silicon photonics platform~\cite{jones2019heterogeneously}. 

A similar wafer bonding process was used in this work to integrate electrically pumped optical gain with ultra-low loss Si$_3$N$_4$ resonators~\cite{xiang20233d}. However, laser integration comes with tradeoffs; previous work showed that processing such as O$_2$ plasma and UV ozone cleaning can increase the loss of Si$_3$N$_4$ waveguides operating at visible wavelengths by UV radiation exciting carriers into defect states, which then become optically absorbing~\cite{neutens2018mitigation}. Furthermore, earlier work on Si$_3$N$_4$ charge-trapping memory devices reported potential excitation through not only UV, but also ion, electron, and X-ray radiation~\cite{naich2004exoelectron}, all of which are possible in the reported fabrication process here. 

In this work, intrinsic $Q$ ($Q_i$) greater than 100 million is demonstrated within a heterogeneous laser process for the first time. The $Q_i$ of ultra-low loss Si$_3$N$_4$ resonators was characterized throughout laser backend processing steps, showing degradation with etches, metal liftoffs, and ion implant, and controllable recovery back up to over 100 million with annealing. 
Previous works have dedicated much study to understanding scattering and absorption-limits in high-$Q$ Si$_3$N$_4$ resonators~\cite{gao2022probing,pfeiffer2018ultra,puckett2021422}; however, no work has shown influences of $Q$ degradation from heterogeneous laser integration. 
Previous heterogeneous laser work has mitigated the effects of laser fabrication by burying the ultra-low-loss Si$_3$N$_4$ layer over 4~$\mu$m below the surface of the laser chips, resulting in $Q_i$ near 40 million (0.5~dB/m) at wavelengths around 1550~nm~\cite{xiang20233d}. To better understand how these $Q$s can be improved to the state-of-the-art values of over 200 million~\cite{jin2021hertz}, the process on this heterogeneous laser platform is repeated with $Q$ measurements performed between laser backend steps, highlighting $Q$ degradation or recovery contributions from individual steps. 

Countless groups across the world have demonstrated powerful functionalities in low-loss Si$_3$N$_4$ with off-chip lasers; the results reported here bolster the credibility of those previous reports promising heterogeneous laser compatibility. With these advancements, laser-integrated low-loss Si$_3$N$_4$ photonics is uniquely positioned to address applications such as field-deployable atom-based clocks and quantum sensors, as well as low-noise microwave photonic oscillators. Due the significant reduction in cost, size, weight, and power compared to traditionally used bulk-optics constrained to lab environments, these laser-integrated Si$_3$N$_4$ photonic solutions can now target emerging markets in GPS-free position, navigation, and timing (PNT), as well as next-generation radar and commercial 5G. 


\begin{figure*}[h]
\centering
\includegraphics[width=7in]{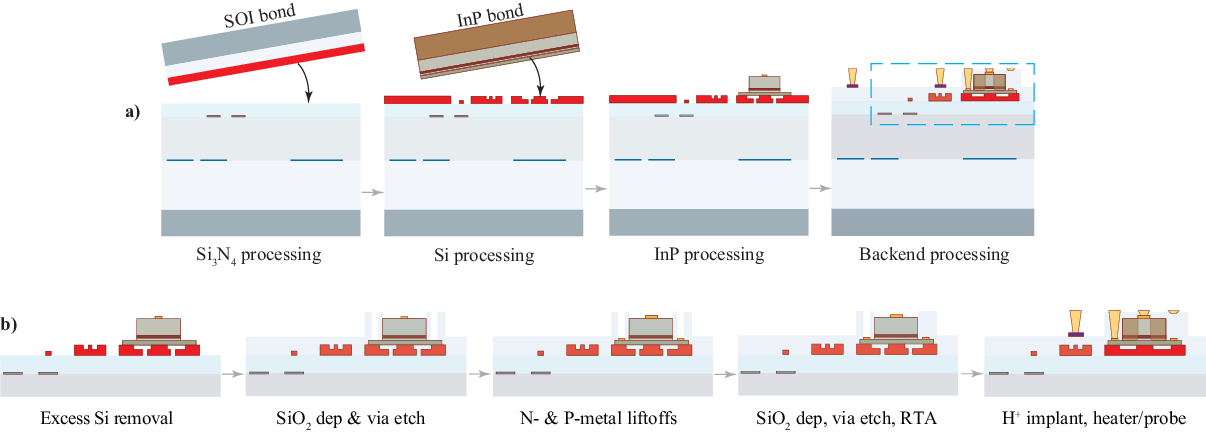}
\caption{\textbf{a)} Simplifed fabrication process for heterogeneously integrating InP/Si lasers on multi-layer Si$_3$N$_4$~\cite{xiang20233d}. The ultra-low-loss, bottom-layer Si$_3$N$_4$ was processed, clad, annealed at a commercial foundry, while the top layer was processed and clad at UCSB. The bottom layer Si$_3$N$_4$ was isolated by over 4~$\mu$m from the surface in the remaining fabrication steps. SOI was bonded and processed to form Si waveguides and grating couplers. InP gain dies were bonded and laser mesas are etched. The boxed section of the backend steps are detailed in \textbf{b)} 
including excess bonded Si removal, SiO$_2$ claddings, via etches, metal liftoffs, hydrogen (H$^{+}$) implantation, and rapid thermal anneal (RTA) steps.
$Q_i$ was measured through backend steps and plotted in Fig.~\ref{fig:Q}.}
\label{fig:process}
\end{figure*}

The heterogeneous laser fabrication is composed of four stages: 1) Si$_3$N$_4$ processing, 2) Si processing, 3) InP processing, and 4) backend processing (Fig.~\ref{fig:process}a). The $Q$ is measured throughout the backend processing steps, further illustrated in Fig.~\ref{fig:process}b.
In the Si$_3$N$_4$ processing stage, the bottom layer 100~nm Si$_3$N$_4$ is deposited by LPCVD on 200~mm wafers at a commercial foundry and annealed over 1000~$^\circ C$ to drive out residual hydrogen and reduce absorption losses~\cite{jin2021hertz}. SiO$_2$ cladding deposition and CMP are also performed at the foundry. After performing another layer of 100~nm LPCVD Si$_3$N$_4$ and coring the wafers into 100~mm wafers through external vendors, the Si$_3$N$_4$ is etched at UCSB, clad with SiO$_2$, and planarized with CMP. Notably, the top layer of Si$_3$N$_4$ is vertically separated from the bottom layer Si$_3$N$_4$ by over 4~$\mu$m SiO$_2$, which helps to protect the bottom layer from contaminants introduced by further laser processing~\cite{xiang20233d}. 
In the Si processing stage, SOI is first bonded, then the Si substrate and buried oxide are subsequently removed by mechanical polishing, Si Bosch etch, and BHF wet etch. The remaining 500~nm Si is then etched into waveguides, grating couplers, and outgassing channels. 
In the InP processing stage, InP gain dies are bonded. The InP substrate is removed with mechanical polishing and HCl wet etch, then the InP laser mesa (P-InP, QW, and N-InP) is etched using a combination of methane/hydrogen/argon (MHA) dry etching and H$_3$PO$_4$-based wet etching~\cite{xiang20233d}.

The backend laser processing steps consist of 1) excess Si removal, 2) SiO$_2$ deposition and via etch, 3) N-metal liftoff, 4) P-metal liftoff, 5) cladding SiO$_2$ deposition, 6) via etch, 7) rapid thermal anneal (RTA), 8) hydrogen (H$^+$) implant, 9) heater metal liftoff, and 10) probe metal liftoff. 
During the InP steps, excess bonded Si is left on top to protect the Si$_3$N$_4$ resonators from process contamination, so in the backend steps, the excess bonded Si is first removed with an unbiased XeF$_2$ dry etch. The optical mode in the Si waveguides is then able to couple into the Si$_3$N$_4$ waveguides without leaking into the excess bonded Si, thus the $Q$ in the lower-layer Si$_3$N$_4$ resonators can be measured via adiabatic tapers from the Si to the Si$_3$N$_4$. Since grating couplers are etched in the Si layer, the $Q$ can be measured by vertical fiber coupling throughout the remaining process steps without dicing the wafer and edge-coupling. 
After excess Si removal, the laser mesa is clad with SiO$_2$ and dry etched with a CF$_4$/O$_2$ chemistry to form vias. P- and N- metal contacts are then formed by electron-beam evaporation and liftoff. Further SiO$_2$ is deposited and vias are etched; RTA is then performed to lower the resistances of the P- and N-contacts. The wafer is then hydrogen implanted to insulate the sides of the laser mesa and confine injected carriers. Lastly, heater and probe metals are deposited for thermo-optic phase tuners and electrical pads.

The $Q$ was measured throughout the laser backend processing steps by sweeping a fiber-MZI-calibrated laser through the resonances of Si$_3$N$_4$ ring resonators with 10~$\mu$m-wide resonator waveguides and 20 GHz FSR. All resonances were measured in the wavelength range of 1540~-~1560~nm, corresponding to the bandwidth of the Si grating couplers. The transmission spectra were fit with a coupled-mode-theory model accounting for backscattering~\cite{pfeiffer2018ultra}, and the $Q$ was extracted from the fitted FWHM of the doublet resonances. Coupling $Q$ was also extracted to confirm that only intrinsic $Q$ varied.

\begin{figure*}[htb]
\centering
\includegraphics[width=7in]{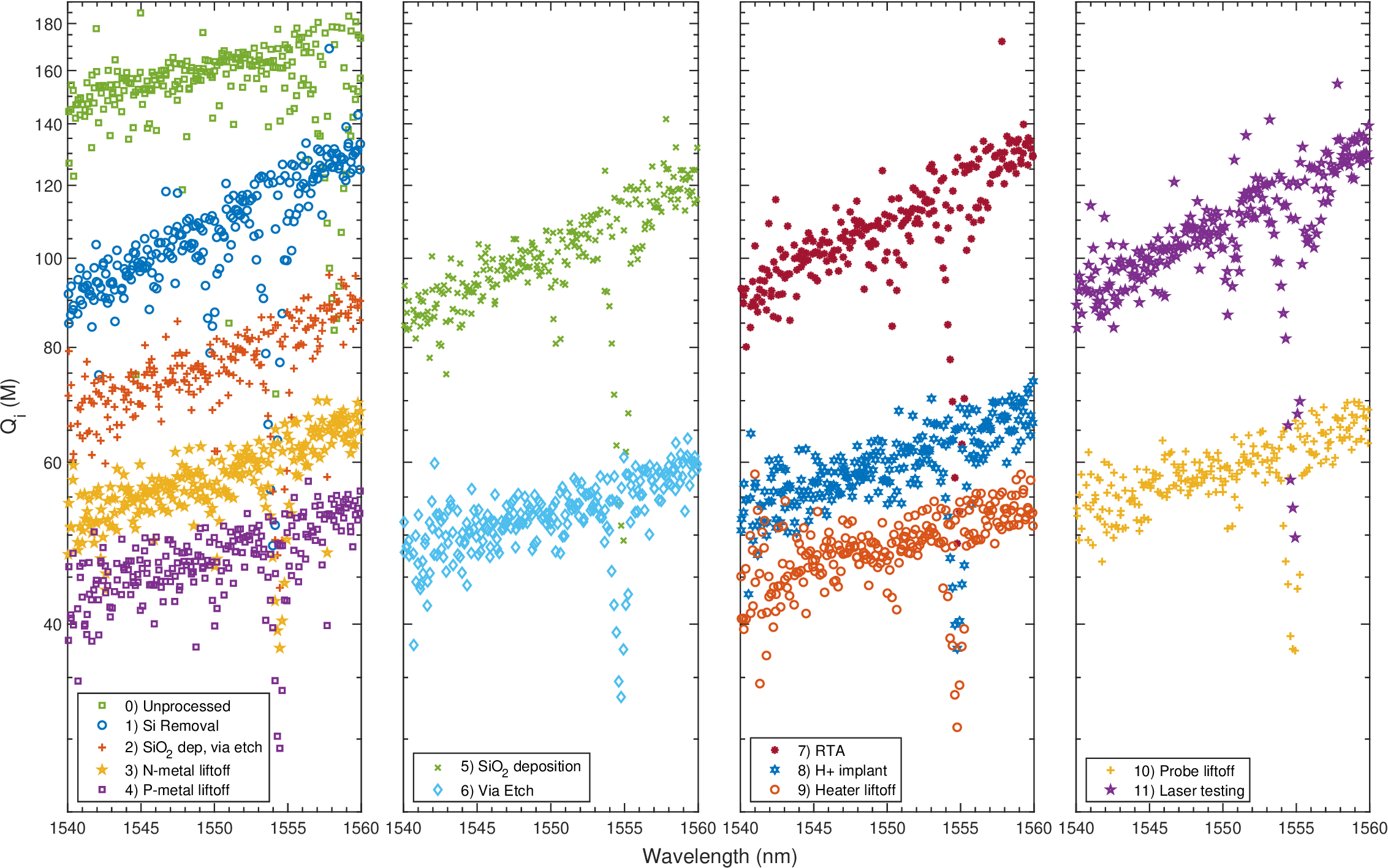}
\caption{$Q_i$ measured before, throughout, and after backend steps consisting of 1) excess Si removal, 2) SiO$_2$ deposition \& via etch, 3) N-metal liftoff, 4) P-metal liftoff, 5) SiO$_2$ deposition, 6) via etch, 7) rapid thermal anneal, 8) hydrogen implant, 9) heater metal liftoff, and 10) probe metal liftoff. 
$Q_i$ measured on the same ring design on another wafer from the same foundry LPCVD Si$_3$N$_4$ batch but without laser processing is plotted as step 0), showing approximate values before laser processing. Measurements after laser testing several months following fabrication (without additional annealing) are plotted as step 11). }
\label{fig:Q}
\end{figure*}


Extracted $Q_i$ is plotted after each backend processing step in Fig~\ref{fig:Q}. After the excess bonded Si is removed (step 1), $Q_i$ over 100 million around 1550~nm is measured, corresponding to about 0.25~dB/m. The $Q_i$ drops down to 75~million with laser mesa cladding and SiO$_2$ via etch (step 2).
The $Q_i$ continues to fall to 50 million with electron-beam evaporation and liftoff of N- and P-contact metals (steps 3 and 4). 
After further SiO$_2$ cladding deposition (step 5), the $Q_i$ is restored above 100 million. The SiO$_2$ ICP-PECVD was performed at 250~$^\circ C$ for 20 minutes, effectively acting as a high-temperature anneal. As previously mentioned, a similar anneal was shown to restore low-loss Si$_3$N$_4$ waveguides at visible wavelengths~\cite{neutens2018mitigation}, and the measurement here corroborates this near 1550~nm.
Once again, $Q_i$ drops with the same SiO$_2$ via etch as previous (CF$_4$ + O$_2$), down to near 50 million at 1550 nm (step 6).
Next, the $Q_i$ recovers with a 30 second, 370 ~$^\circ C$ rapid thermal anneal (step 7) - again, similar to previous work although for a significantly shorter time~\cite{neutens2018mitigation}. 
In the last steps of the backend process, the $Q_i$ continues to degrade with hydrogen (H$^{+}$) implant (step 8) and heater metal liftoff (step 9) down to seemingly saturated values near 50 million. Details of the hydrogen implant are provided in previous works~\cite{davenport2017heterogeneous}. 
The $Q_i$ recovers slightly from probe metal deposition (step 10), most likely related to heating from the thick metal stack. 
To mitigate risk associated with migrating the H$^{+}$ dopants, further annealing was held off until after laser testing. However, upon measuring after months of laser testing, the $Q_i$ was found to have recovered above 100 million (step 11). Thus, $Q_i$ over 100 million was found to be recoverable immediately by high temperature annealing and also gradually without annealing.

\begin{figure}[htb]
\centering
\includegraphics[width=\linewidth]{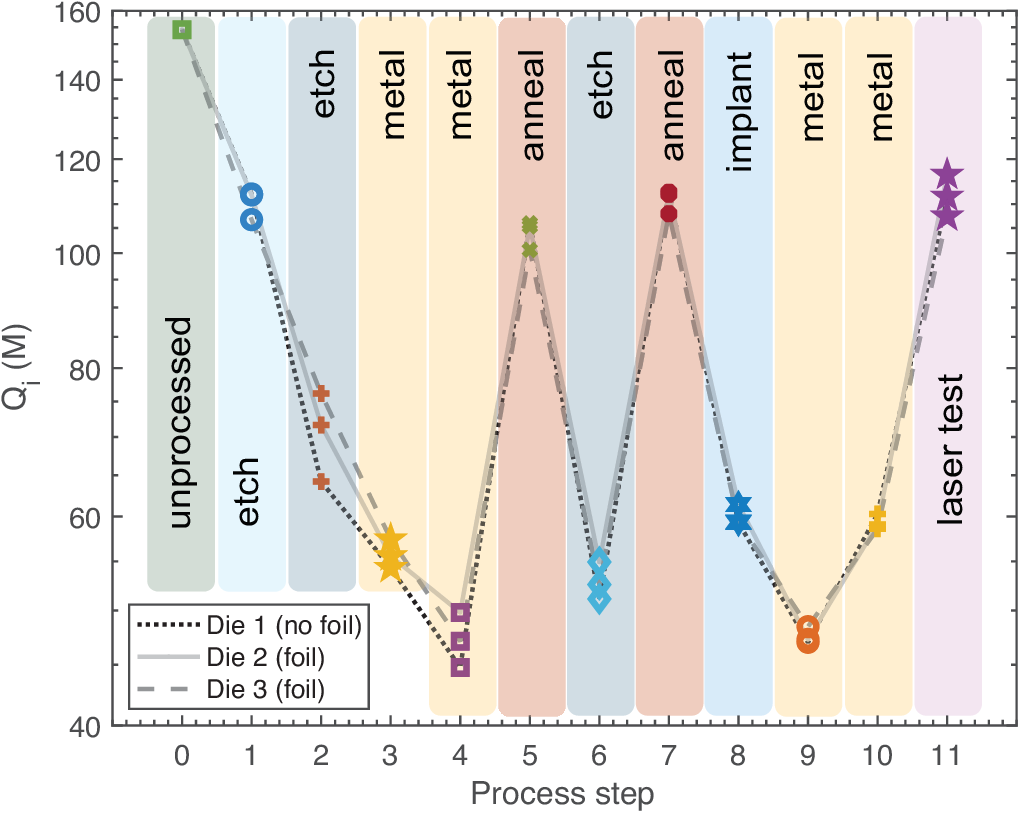}
\caption{Wavelength averages of $Q_i$ with process steps across three dies from the same wafer. Dies 2 and 3 were protected with aluminum foil during metal liftoff steps, while die 1 was not. $Q_i$ over 100~million was measured after Si removal (backend step 1). The $Q_i$ generally dropped with dry etches, metal liftoffs, and H$^{+}$ implant and recovered with high temperature anneals. Over time, the $Q_i$ recovered to its initially measured values, measured after several months of laser testing. Again, measured Q data from a wafer without laser processing is included as step 0) as a reference.  SiO$_2$ etches (steps 2 and 6), anneals (steps 5 and 7), and metal liftoffs (steps 3, 4, 9, 10) are grouped together by color since they are similar processes. Note that the XeF$_2$ etch in step 1 is unbiased and categorized differently than the biased SiO$_2$ etches.}
\label{fig:averages}
\end{figure}

A summary of the drops and recoveries in $Q_i$ is plotted in Fig.~\ref{fig:averages} by averaging the data over wavelength and plotting with process step. Data from two other dies on the same wafer processed simultaneously are also included to show the same trends across a wafer. The measured Si$_3$N$_4$ resonators on dies 2 and 3 were protected with aluminum foil during metal liftoff steps, while the measured resonators on die 1 were not protected, showing negligible differences from foil protection. Significant $Q_i$ drops are correlated with etches, implant, and metal liftoffs with varying strengths, while recoveries are correlated with anneals. The $Q_i$ measured from another wafer from the same foundry Si$_3$N$_4$ LPCVD batch but without further laser processing is also plotted as step 0) on both Figures~\ref{fig:Q} and \ref{fig:averages} approximating the $Q_i$ values before laser processing.

In conclusion, the intrinsic $Q$ of Si$_3$N$_4$ resonators was characterized throughout heterogeneous laser processing steps, dropping with dry etches, ion implant, and metal liftoffs, while recovering with anneals and thermal relaxation over a long time constant. Recovery values saturate near 110 million and degradation values saturate near 50 million, both at 1550 nm. All $Q$ measurements were taken within a day of the associated processing step, after stripping the residual photoresist with solvent and O$_2$ plasma cleans when necessary (i.e. SiO$_2$ deposition and RTA did not require photoresist). As a result, the time constant for recovery without annealing is greater than days. Previous work showed that the mechanism of increasing loss was the population of optically absorbing defect states in Si$_3$N$_4$ by UV excitation only~\cite{neutens2018mitigation}, while the work presented here introduces many more potential excitation mechanisms due to the variety of processing necessary for laser fabrication. As previously mentioned, Si$_x$N$_y$ has a high concentration of defect states, which can be populated not only by UV, but also by electron, ion, and UV radiation~\cite{naich2004exoelectron}. All of these are potentially introduced in the fabrication process reported here;
for example, electron-beam evaporation of metals is known to eject X-rays, ionized metal vapor, and secondary electrons~\cite{volmer2021solve,mayo1976radiation}. 
Aluminum foil protection covering the resonators was used to test mitigation of Q degradation in all metal evaporation steps with negligible differences found, as shown in Fig~\ref{fig:averages}; however, the foil also makes penetration of UV and electron radiation into the Si$_3$N$_4$ unlikely.
Future work dedicated to separating the effects of UV, ion, electron, and X-ray radiation should start by measuring $Q$ degradation from O$_2$ plasma cleans of increasing times, since this was common to nearly all steps in varying degrees depending on residual photoresist. X-ray radiation can perhaps be tested with electron beam evaporation systems at different beam currents necessary for evaporation of the various metals used in the P- and N-contact metal stacks. 
Considering the higher $Q$s of 160 million measured on another wafer from the same Si$_3$N$_4$ foundry run but without heterogeneous laser processing, further prevention before the backend steps may be possible and will need to be investigated. 
Lastly, the same suspected Si$_3$N$_4$ defect states responsible for the $Q$ fluctuations in this work have also been postulated to be involved in the coherent photogalvanic effect, resulting in large effective $\chi^{(2)}$ nonlinearities in Si$_3$N$_4$~\cite{billat2017large}; thus, further investigation to link these two is necessary. 

Despite the many potential degradation mechanisms, Si$_3$N$_4$ resonators with intrinsic $Q$s greater than 100 million were demonstrated after full heterogeneous InP/Si laser integration. Factoring in the powerful functionalities of microresonator frequency combs, second harmonic generation, and extreme laser noise reduction, this laser-integration Si$_3$N$_4$ photonics platform paves the way for scalable manufacturing and field-deployment of next-generation atomic clocks, quantum sensors, and low-noise microwave oscillators.

\begin{backmatter}
\bmsection{Funding}
This research was supported by DARPA GRYPHON (HR0011-22-2-009)

\bmsection{Acknowledgments}
The authors thank Minh Tran, Demis John and Brian Thibeault for helpful discussions. A portion of this work was performed in the UCSB Nanofabrication Facility, an open access laboratory.

\bmsection{Disclosures}
The authors declare no conflicts of interest.

\bmsection{Data Availability Statement}
Data underlying the results presented in this paper are not publicly available at this time but may be obtained from the authors upon reasonable request.

\end{backmatter}

\bibliography{references}

\end{document}